\documentclass[aip,prx,amsmath,amssymb]{revtex4-1}
\usepackage{amsmath}
\usepackage{amssymb}
\usepackage{hyperref}
\usepackage{graphicx}
\usepackage{multirow}
\usepackage{caption}
\usepackage{subcaption}
\usepackage{color}
\usepackage{amsthm,enumitem}

\begin{document}

This manuscript has been authored by UT-Battelle, LLC, under Contract No. DE6AC05-00OR22725
with the U.S. Department of Energy. The U.S. Government is authorized to reproduce and distribute reprints
for Government purposes notwithstanding any copyright notation hereon. The Department of Energy will
provide public access to these results of federally sponsored research in accordance with the DOE Public
Access Plan (http://energy.gov/downloads/doe-publicaccess-plan).
\mbox{}
\thispagestyle{empty}
\newpage 

\title{Modeling Kinetic Effects of Charged Vacancies on Electromechanical Responses of Ferroelectrics: Rayleighian Approach}
\author{Rajeev Kumar}
\email{kumarr@ornl.gov}
\affiliation{Center for Nanophase Materials Sciences, Oak Ridge National Laboratory, Oak Ridge, TN-37831}

\author{Shuaifang Zhang}
\affiliation{Center for Nanophase Materials Sciences, Oak Ridge National Laboratory, Oak Ridge, TN-37831}

\author{P. Ganesh}
\email{ganeshp@ornl.gov}
\affiliation{Center for Nanophase Materials Sciences, Oak Ridge National Laboratory, Oak Ridge, TN-37831}

\date{\today}
\begin{abstract}
Understanding time-dependent effects of charged vacancies on electromechanical responses of materials is at the forefront of research for designing materials exhibiting metal-insulator transition, and memresistive behavior.  A Rayleighian approach is used to develop a model for studying the non-linear kinetics of reaction leading to generation of vacancies and electrons by the dissociation of vacancy-electron pairs. Also, diffusion and elastic effects of charged vacancies are considered to model polarization-electric potential and strain-electric potential hysteresis loops. The model captures multi-physics phenomena by introducing couplings among polarization, electric potential, stress, strain, and concentrations of charged (multivalent) vacancies and electrons (treated as classical negatively charged particles), where the concentrations can vary due to association-dissociation reactions. Derivation of coupled time-dependent equations based on the Rayleighian approach is presented. Three limiting cases of the governing equations are considered highlighting effects of 1) non-linear reaction kinetics on the generation of charged vacancies and electrons, 2) the Vegard's law (i.e., the concentration-dependent local strain) on asymmetric strain-electric potential relations, and 3) coupling between a fast component and the slow component of the net polarization on the polarization-electric field relations. The Rayleighian approach discussed in this work should pave the way for developing a multi-scale modeling framework in a thermodynamically consistent manner while capturing multi-physics phenomena in ferroelectric materials.    
\end{abstract}

\maketitle
\section{Introduction}
Fundamental principles, which can help in design of materials exhibiting metal-insulator transition\cite{mit_review} and memory effects, are under extensive scrutiny by a number of researchers in the fields of spintronics\cite{awschalom2007challenges}, ferroelectrics\cite{rabe2007modern}, and neuromorphic computing\cite{fe_neuro}. Experimental discoveries\cite{parkin_science} connecting oxygen vacancies to unconventional ferroelectricity in films and metal-insulator transitions have led to research activities focused on understanding effects of the vacancies on material properties. These activities include experimental measurements, which can decouple contributions to switching of polarization from polar crystal phases and vacancy migration under applied electric fields, transient effects of vacancy ordering\cite{kelley2022oxygen}, and electrochemical effects due to surfaces\cite{yang2017mixed}. In parallel, models capturing effects of oxygen vacancies at different levels of approximations have been developed, which include molecular models based on density functional theory\cite{kelley2022oxygen}, reactive force fields\cite{akbarian2019understanding}, and continuum models\cite{xiao2008continuum} capturing long-wavelength physics. These research activities, experimental and theoretical, have unequivocally established that electrostatic effects due to vacancies are of paramount importance in controlling properties of materials at length and time scales relevant to application of materials in microelectronics. For example, Lin \textit{et al.}\cite{lin2020oxygen} proposed a hypothesis  that enhanced remnant polarization of a composite containing ferroelectric barium titante (BaTiO$_3$) and metallic non-ferroelectric oxide of SrRuO$_3$ may be attributed to the accumulation of oxygen vacancies at the BaTiO$_3$/SrRuO$_3$ interface. Similarly, thin films ($\sim 80$ nm) of  $\mbox{BaTiO}_3$ grown on $\mbox{SrRuO}_3//\mbox{SrTiO}_3$ have been studied recently\cite{kelley2022oxygen} to understand effects of oxygen vacancy injection on electromechanical responses of the films. It has been established\cite{kelley2022oxygen} that enhanced electromechanical responses in these films can be sustained by injection of oxygen vacancies and kinetic effects due to the vacancies were postulated to be responsible for the responses.

Modeling kinetic effects due to vacancies in thin films of ferroelectric materials requires use of a simulation method, which can capture effects of multiple phases (varying in polarization and crystal symmetry) and multi-physics phenomena due to couplings among polarization, electrostatic potential, strain, and vacancies. For long length and time scales, phase field method has been used to model ferroelectric\cite{li2001phase,saha2019phase,wang2019understanding} materials, using parameters, which can be either inferred from experimental data or estimated using density functional theory (DFT). The phase field method has been used to model phenomena occurring at long time scales (of the order of seconds), such as domain wall motion \cite{li2001phase} and domain nucleation\cite{saha2019phase}. The method incorporates a thermodynamic free energy similar to the originally proposed by Landau, Ginzburg, and Devonshire for ferroelectricity in the bulk (i.e., without surfaces), generalized to inhomogeneous systems and time domain. The free energy is typically written in terms of a space and time dependent polarization vector, and its gradients. Coefficients of various terms in the free energy can be either kept phenomenological or estimated from first principles calculations such as DFT after identifying underlying origins of polarization in terms of atomic displacements . For example, the free energies for some ferroelectrics materials are summarized by Chen \cite{chen2007appendix}.

In the last two decades, a number of phase field models have been developed for understanding the effects of vacancies in ferroelectric materials. For example, Zhang \textit{et al.} \cite{zhang2010oxygen} developed a phase field model for ferroelectrics with oxygen vacancies by treating them as defect dipoles and investigated the oxygen-vacancy-induced memory effect and large recoverable strain in BaTiO$_3$. Cao \textit{et al.} \cite{cao2014effect} treated the vacancies as charged defects and developed a phase field method for BaTiO$_3$ by coupling time dependent equations for polarization with the Poisson-Nernst-Planck equations for density of charges. Shindo \textit{et al.}\cite{shindo2015phase} used a phase field model to study the electromechanical response of polycrystal BaTiO$_3$ with oxygen vacancies by treating the vacancies as defect dipoles. Recently, Fedeli \textit{et al.}\cite{fedeli2019phase} presented a phase field model for ferroelectrics by treating defects in the single crystal and polycrystal structures as voids, charged point defect and polarization pinned objects, such that the polarization pinned objects retain their polarization during the cycling of electric field. Lastly, Kelley \textit{et al.} \cite{kelley2022oxygen} proposed a new free energy by including the impact of oxygen vacancies on not only polarization, but also on strains, which they used to understand asymmetry in the polarization-electrostatic potential ($P-\psi$) loops with different concentrations of oxygen vacancies. In order to simply the semi-analytical analysis, they invoked an assumption that the vacancies move much faster than the polarization so that a steady state approximation for relaxation of the densities of the vacancies can be justified. Despite these modeling efforts spanning two decades related to the modeling of vacancies and ferroelectrics, a unified theoretical framework, which can be used to construct thermodynamically consistent kinetic models for thin films of ferroelectric materials with vacancies is still lacking. Development of such a theoretical framework is the main goal of this paper. Limiting cases are discussed to highlight key features of this framework.  


In this manuscript, a Rayleighian approach is used to develop time and space dependent equations for polarization, densities of vacancies, electrons, and their pairs, electrostatic potential, and strains. The approach has its basis in principles of linear irreversible thermodynamics and ensures that the second law of thermodynamics is obeyed. Numerical solutions of coupled set of equations will be presented in another manuscript for the purpose of constructing average polarization-electric potential and strain-electric potential hysteresis loops. The manuscript is organized as follows: details of energy functional, the Rayleighian, and the coupled equations are presented in the section ~\ref{sec:model}. Different limits of the coupled set of equations are analyzed in section ~\ref{sec:results}. Conclusions are presented in section ~\ref{sec:conclusions}.

\section{Model building for thin films of ferroelectrics using the Rayleighian approach}~\label{sec:model}
A set of time-dependent equations can be derived for simulating ferroelectric materials like $\mbox{BaTiO}_{3}$ with vacancies. These equations can be derived using principles of irreversible thermodynamics. For the derivation, we decompose local polarization into fast and slow components resulting from electronic and atomic/orientational motions, respectively. The fast component of the polarization ($=\mathbf{P}_e(\mathbf{r},t)$) at a location $\mathbf{r}$ at time $t$ is assumed to be in equilibrium with the electric field ($\mathbf{E}(\mathbf{r},t) = -\nabla \psi(\mathbf{r},t)$) i.e., $\mathbf{P}_e(\mathbf{r},t) = \epsilon_0 \epsilon_{\infty} \mathbf{E}(\mathbf{r},t)$ so that $\epsilon_0$ is the permittivity of vacuum, $\epsilon_{\infty}$ is the infinite frequency dielectric constant of the material, and $\psi$ is the electrostatic potential. In contrast, the slow component ($=\mathbf{P}(\mathbf{r},t)$) can be out-of-equilibrium and time-dependent equations are derived for this component. Such a decomposition of the polarization is similar to the work by Marcus\cite{marcus_1} focused on electron transfer processes. Coupling of the slow component of polarization with the fast component is considered by introducing lattice strain. Diffusion of charged vacancies and electrons are considered by using their local densities as additional order parameters, and following our previous work related to ion transport in polymerized ionic liquids\cite{kumar2017}. Furthermore, reaction kinetics leading to electrochemical effects, which result from charge generation and recombination of positively charged vacancies with electrons, are considered. Strain and concentration of the vacancies are coupled by the Vegard's law\cite{ashcroftPRA}, which is an empirical rule leading to a linear relation between the strain and the concentration of the vacancies. Mathematical framework leading to equations, which can introduce couplings among polarization, strain, diffusion, and reaction kinetics, is presented below. 

Any material perturbed from equilibrium by the application of a small external force will try to approach equilibrium through various dissipative processes. Dissipative processes and the path must lead to the positive rate of entropy production as per the second law of thermodynamics\cite{onsager1,groot1962non,kumar2017}.  These statements were cast into a rigorous mathematical framework by Onsager\cite{onsager1} after defining a functional called the Rayleighian, which consists of the rate of change of ``free'' energy (or entropy for isothermal processes) and a dissipation function. According to Onsager's variational principle, the true dynamics of a system is one that demonstrates the least dissipation. Linear relations among various fluxes and forces can hence be derived by minimizing the Rayleighian with respect to the velocities leading to the maximum rate of entropy production and the resulting equations define the most probable path towards an equilibrium\cite{machulp1953,doi2019}. Details of this so-called Onsager variational principle are presented elsewhere\cite{onsager1,machulp1953,Doi_2011,doi2019}. Construction of the Rayleighian requires identification of independent fluxes, reaction rates, and constraints, which are discussed in the next subsection.

\subsection{Independent Fluxes, Rates, and Constraints}
A model for understanding effects of vacancies can be developed using the principles of irreversible thermodynamics\cite{onsager1,groot1962non,kumar2017}. In particular, we use Onsager's variational principle\cite{Doi_2011} to derive a set of equations for the polarization ($\boldsymbol{P}(\mathbf{r},t)$), strain tensor ($\boldsymbol{\varepsilon}(\mathbf{r},t)$), number densities of electroactive vacancies ($\rho_{+}(\mathbf{r},t)$), electrons ($\rho_{-}(\mathbf{r},t)$), and their pairs ($\rho_{\pm}(\mathbf{r},t)$), which are coupled via the electric field ($\boldsymbol{E}(\mathbf{r},t)$). The number densities, $\rho_{+}(\mathbf{r},t)$, $\rho_{-}(\mathbf{r},t)$, and $\rho_{\pm}(\mathbf{r},t)$ are assumed to satisfy
\begin{eqnarray}
    \frac{\partial \rho_{i}(\mathbf{r},t)}{\partial t}  &=& -\nabla \cdot \mathbf{j}_i (\mathbf{r},t) + S_i(\mathbf{r},t) \quad \mbox{for}\quad i = +,-,\pm\label{eq:den} 
\end{eqnarray}
where $\mathbf{j}_i (\mathbf{r},t) = \rho_i (\mathbf{r},t) \mathbf{ v}_i (\mathbf{r},t)$ is the diffusive flux so that $\mathbf{v}_i (\mathbf{r},t)$ is the collective velocity of $i=+,-,\pm$. In Eq. ~\ref{eq:den}, $S_i (\mathbf{r},t)$ is the rate of change of the number density of $i$ resulting from the dissociation and the formation of vacancy-electron pairs. We assume that the number densities satisfy the no-void condition at all locations at all times so that the sum of the volume fractions defined as $\phi_i (\mathbf{r},t) = \rho_i (\mathbf{r},t)/\rho_{io}$ is unity i.e.,
\begin{eqnarray}
    \sum_{i=+,-,\pm}\phi_{i}(\mathbf{r},t)  &=& 1\label{eq:incompress} 
\end{eqnarray}
where $1/\rho_{io}$ is the molar volume of $i$. Using Eq. ~\ref{eq:den} and ~\ref{eq:incompress}, a constraint on the 
fluxes and the reaction rates is obtained, which is written as
\begin{eqnarray}
    \sum_{i=+,-,\pm}\left[-\nabla \cdot\left\{ \frac{\mathbf{j}_i (\mathbf{r},t)}{\rho_{io}}\right\} + \frac{S_i(\mathbf{r},t)}{\rho_{io}}\right] &=& 0 \label{eq:fluxconstraint} 
\end{eqnarray}
Now, we need to construct relations between $\mathbf{j}_i (\mathbf{r},t)$, $S_i (\mathbf{r},t)$ and thermodynamic forces. In the following, this is accomplished by considering a functional called the Rayleighian ($R$). Before defining the Rayleighian, we need to identify relevant indepedent fluxes ($\mathbf{j}_i$), rates ($S_i$), and thermodynamic forces. 

For identifying independent fluxes ($\mathbf{j}_i$), rates ($S_i$), and other constraints, we consider the rate of change of the total number ($=M(t)$) of particles in the vacancies, the electrons, and the pairs, defined as $M(t) = \int d\mathbf{r}\,\left[\rho_+(\mathbf{r},t) + \rho_-(\mathbf{r},t) + (z_+ + 1)\rho_{\pm}(\mathbf{r},t)\right]$. Due to the fixed number of particles at all times, the rate of change of $M(t)$ must be zero and the rate is given by
\begin{eqnarray}  
\frac{d M(t)}{d t} &=& \int d\mathbf{r}\, \left [-\nabla \cdot \left\{\mathbf{j}_+(\mathbf{r},t) + \mathbf{j}_-(\mathbf{r},t) + (z_+ + 1)\mathbf{j}_{\pm}(\mathbf{r},t)\right\} \right . \nonumber \\
&& \left. + \left\{S_+(\mathbf{r},t) + S_-(\mathbf{r},t) + (z_+ + 1)S_{\pm}(\mathbf{r},t)\right\}\right] \nonumber \\
&=& \int d\mathbf{\Gamma}\, \mathbf{\hat{n}}\cdot \left\{\mathbf{j}_+(\mathbf{r},t) + \mathbf{j}_-(\mathbf{r},t) + (z_+ + 1)\mathbf{j}_{\pm}(\mathbf{r},t)\right\} \label{eq:rateM}
\end{eqnarray}
where we have used Eqs. ~\ref{eq:den} and assumed 
$S_+(\mathbf{r},t) + S_-(\mathbf{r},t) + (z_+ + 1)S_{\pm}(\mathbf{r},t) = 0$ so that the total number of particles remains the same due to the dissociation-association reactions. Here, $\Gamma$ represents the surface enclosing the volume under consideration  so that $\mathbf{\hat{n}}$ is an outward normal at the surface and we have used the divergence theorem. Now, equating $d M(t)/d t = 0$, Eq. ~\ref{eq:rateM} is satisfied if 
\begin{eqnarray}  
\mathbf{\hat{n}}\cdot \left\{\mathbf{j}_+(\mathbf{r},t) + \mathbf{j}_-(\mathbf{r},t) + (z_+ + 1)\mathbf{j}_{\pm}(\mathbf{r},t)\right\} &=& 0 \label{eq:boundaryfluxconstraint}
\end{eqnarray}
at the surface. Another constraint for the boundary fluxes is obtained by considering the rate of change of the total charge ($=C(t)$) of the vacancies, and the electrons, defined as $C(t) = e\int d\mathbf{r}\left[z_+\rho_+(\mathbf{r},t) - \rho_-(\mathbf{r},t)\right]$, which must be zero due to the global electroneutrality at all times. The rate of change of $C(t)$ is given by
\begin{eqnarray}  
\frac{d C(t)}{d t} &=& e\int d\mathbf{r}\, \left [-\nabla \cdot \left\{z_+ \mathbf{j}_+(\mathbf{r},t) - \mathbf{j}_-(\mathbf{r},t)\right\} + \left\{z_+S_+(\mathbf{r},t) - S_-(\mathbf{r},t)\right\}\right] \nonumber \\
&=& \int d\mathbf{\Gamma}\, \mathbf{\hat{n}}\cdot \left\{z_+\mathbf{j}_+(\mathbf{r},t) - \mathbf{j}_-(\mathbf{r},t)\right\}  = 0 \label{eq:rateC}
\end{eqnarray}
where we have assumed that $z_+S_+(\mathbf{r},t) - S_-(\mathbf{r},t) = 0$ i.e., the rate of charge generation due to the dissociation-association reactions is taken to be zero. Combining $S_+(\mathbf{r},t) + S_-(\mathbf{r},t) + (z_+ + 1)S_{\pm}(\mathbf{r},t) = 0$ with $z_+S_+(\mathbf{r},t) - S_-(\mathbf{r},t) = 0$ leads to 
\begin{eqnarray}
    S_+(\mathbf{r},t) &=& \frac{1}{z_+} S_-(\mathbf{r},t) = - S_{\pm}(\mathbf{r},t) \equiv - S(\mathbf{r},t) \label{eq:rateconstraint} 
\end{eqnarray}
for the association-dissociation reactions involving  electroneutral vacancy-electron pairs, positively charged vacancies with charge  $z_+ e$ (so that $e$ is the charge of an electron) and neutralizing electrons. For example, $z_+ = 2$ in the case of oxygen vacancies. Eq. ~\ref{eq:rateconstraint} implies that there is only one independent reaction rate, which is taken to be $S(\mathbf{r},t)$ and defined by Eq. ~\ref{eq:rateconstraint}. Now, Eqs. ~\ref{eq:fluxconstraint} and ~\ref{eq:rateconstraint} imply that there are only two independent fluxes out of the three fluxes, $\mathbf{j}_i$. Based on a molecular description of the diffusion resulting from frictional forces and relative motion of molecules, we work with 
the relative fluxes defined as $\mathbf{\hat{j}}_i (\mathbf{r},t) = \phi_i (\mathbf{r},t) \left[\mathbf{ v}_i (\mathbf{r},t)- \mathbf{ v} (\mathbf{r},t)\right]$, where
\begin{eqnarray} 
    \mathbf{v} (\mathbf{r},t) &=& \sum_{i=+,-,\pm}\phi_{i}(\mathbf{r},t) \mathbf{ v}_i (\mathbf{r},t) \label{eq:barycentric}
\end{eqnarray}
These relative fluxes are related by the relation
\begin{eqnarray}
    \sum_{i=+,-,\pm}\mathbf{\hat{j}}_i (\mathbf{r},t) &=& 0\label{eq:relativefluxconstraint} 
\end{eqnarray}
and the Rayleighian can be written in terms of any two relative fluxes. In here, we choose $\mathbf{\hat{j}}_+$ and $\mathbf{\hat{j}}_-$ to study the effects of diffusion of the vacancies and the electrons, respectively. Eqs. ~\ref{eq:den} are rewritten in terms of these independent fluxes and rates in the form
\begin{eqnarray}
    \frac{\partial \phi_{k}(\mathbf{r},t)}{\partial t}  &=& -\nabla \cdot \mathbf{\hat{j}}_k (\mathbf{r},t) + \frac{1}{\rho_{ko}}S_k(\mathbf{r},t) - \nabla \cdot \left[\mathbf{\phi}_k (\mathbf{r},t)\mathbf{ v} (\mathbf{r},t)\right]\quad \mbox{for}\, k = +,-.\label{eq:denrewritten} 
\end{eqnarray}
and Eq. ~\ref{eq:fluxconstraint} can be written as
\begin{eqnarray}
   \nabla \cdot \mathbf{v} (\mathbf{r},t) + S(\mathbf{r},t)\left\{\frac{1}{\rho_{+o}} + \frac{z_+}{\rho_{-o}}-\frac{1}{\rho_{\pm o}}\right\} &=& 0 \label{eq:fluxconstraintre} 
\end{eqnarray}
Eqs. ~\ref{eq:rateM} and ~\ref{eq:rateC} are rewritten in terms of $\mathbf{\hat{j}}_{i=+,-,\pm}\cdot \mathbf{\hat{n}}$ as 

\begin{eqnarray}  
\mathbf{\hat{n}}\cdot \left\{\rho_{+o}\mathbf{\hat{j}_+}(\mathbf{r},t) + \rho_{-o}\mathbf{\hat{j}_-}(\mathbf{r},t) + (z_+ + 1)\rho_{\pm o}\mathbf{\hat{j}_{\pm}}(\mathbf{r},t)\right\} && \nonumber\\
+ \mathbf{\hat{n}}\cdot \left\{\left[\rho_{+o} \phi_{+}(\mathbf{r},t) + \rho_{-o} \phi_{-}(\mathbf{r},t) + (z_+ + 1)\rho_{\pm o}\phi_{\pm}(\mathbf{r},t)\right]\mathbf{v} (\mathbf{r},t)\right\} &=& 0 \label{eq:rateMre} \\
\mathbf{\hat{n}}\cdot \left\{z_+\rho_{+o}\mathbf{\hat{j}_+}(\mathbf{r},t) - \rho_{-o}\mathbf{\hat{j}_-}(\mathbf{r},t) + \left[z_+\rho_{+o}\phi_{+}(\mathbf{r},t) - \rho_{-o}\phi_{-}(\mathbf{r},t)\right]\mathbf{v}(\mathbf{r},t)\right\}  &=& 0 \label{eq:rateCre}
\end{eqnarray}
These equations show that there are two boundary fluxes, which are independent. These are chosen to be $\mathbf{\hat{j}}_{k=+,-}\cdot \mathbf{\hat{n}}$. In summary, independent fluxes and rates are $\mathbf{\hat{j}}_{k=+,-}, \mathbf{\hat{j}}_{k=+,-}\cdot \mathbf{\hat{n}}$, and $S$, respectively. In addition, we need to consider the constraint written as Eq. ~\ref{eq:fluxconstraintre} and discussed above.

\subsection{Rayleighian}
For applying the Onsager's variational principle, a Rayleighian ($R(t)$) for the thin films of ferroelectric materials can be defined as\cite{kumar2017,cummings2018}  
\begin{eqnarray}
    R(t) &=& \frac{d H(t)}{d t} + W(t) - \int d\mathbf{r}\, p(\mathbf{r},t)\left[\mathbf{\nabla} \cdot \mathbf{v} (\mathbf{r},t) + S(\mathbf{r},t)\left\{\frac{1}{\rho_{+o}} + \frac{z_+}{\rho_{-o}}-\frac{1}{\rho_{\pm o}}\right\}\right] \label{eq:ray} 
\end{eqnarray}
where $H(t)$ is the time-dependent energy functional (presented in the next subsection) and $p(\mathbf{r},t)$ is a Lagrange's multiplier to enforce the constraint (cf. Eq. ~\ref{eq:fluxconstraintre}), and $W$ is the dissipation function. We should point out that the functional form for the dissipation function is assumed to be known in order to use the variational principle and there is no prescription for deriving it. In this paper, we present a functional form for the dissipation function by including various multi-physics phenomena and ensuring that the derived equations lead to known relations in various limiting cases. For example, $W$ has contributions from the coupling of the polarization with the lattice velocity, derived by Hubbard and Onsager\cite{hubbard1977dielectric} using the approximation of fast rotational relaxation\cite{groot1962non,felderhof1999hydrodynamics}. Also, contributions from friction, which result from relative motions of charged vacancies and electrons with respect to the velocity $\mathbf{v}(\mathbf{r},t)$ (cf. Eq. ~\ref{eq:barycentric}), are included\cite{kumar2017,Doi_2011}. Adding these contributions, $W$ can be written as  
\begin{eqnarray}
    W(t) &=& \frac{\tau_p}{2}\int d\mathbf{r}\,\left[\frac{\partial \mathbf{P}(\mathbf{r},t)}{\partial t} + \left[\mathbf{v}_l(\mathbf{r},t)\cdot \nabla\right]\mathbf{P}(\mathbf{r},t) + \frac{1}{2}\mathbf{P}(\mathbf{r},t) \times \left[\nabla \times \mathbf{v}_l(\mathbf{r},t)\right]\right]^2 \nonumber \\
    && + \frac{1}{2} \int d\mathbf{r}\sum_{k=+,-}\sum_{k'=+,-} L_{kk'}(\mathbf{r},t)\rho_{ko}\rho_{k'o}\mathbf{\hat{j}}_{k}(\mathbf{r},t) \cdot \mathbf{\hat{j}}_{k'}(\mathbf{r},t) \nonumber \\
    && + \frac{1}{2} \int d\mathbf{r}\sum_{k=+,-}\sum_{k'=+,-} M_{kk'}(\mathbf{r},t)\rho_{ko}\rho_{k'o}\left[\mathbf{\hat{j}}_{k}(\mathbf{r},t)\cdot \mathbf{\hat{n}}\right] \left[\mathbf{\hat{j}}_{k'}(\mathbf{r},t)\cdot \mathbf{\hat{n}}\right]\nonumber \\
&& + \frac{1}{2} \int d\mathbf{r} \,\omega(\mathbf{r},t) S^2(\mathbf{r},t) \label{eq:dissipation}
\end{eqnarray}
Here, $\tau_p$ and $L_{kk'}$ are parameters characterizing time-scale for change in polarization $\mathbf{P}$, and friction coefficient for motion of the vacancies, and electrons relative to $\mathbf{v}(\mathbf{r},t)$ (defined by Eq. ~\ref{eq:barycentric}). Similarly, $M_{kk'}$ are the parameters characterizing the dissipation due to the relative fluxes at the boundaries. Also, $\mathbf{v}_l = \sum_{l=1,2,3}\left[\frac{\partial u_l(\mathbf{r},t)}{\partial t}\right] \hat{i}_l$ is net velocity of an underlying lattice so that $\hat{i}_l$ are unit vectors and $u_l(\mathbf{r},t)$ is the displacement of underlying atoms at location $\mathbf{r}$ at time t. In general, $L_{kk'}$ and $M_{kk'}$ can be concentration dependent but each matrix with either $L_{kk'}$ or $M_{kk'}$ as its elements must be positive definite for the positive entropy 
production. The last term in Eq. ~\ref{eq:dissipation} is the dissipation due to the reaction with a prefactor $\omega$, which will be related to the rate of vacancy-electron pair dissociation and recombination. 


In the following, we present explicit expression for $H(t)$ by using thermodynamic free energy and by generalizing it after considering additional effects of the charged vacancies. 

\subsection{Energy Functional}
For deriving a set of time-dependent equations, we use a time ($t$)-dependent energy functional, written as 
\begin{eqnarray}
    H(t) &=& \int d\mathbf{r} \,\left[ H_{LGD}\left\{\boldsymbol{P}\right\} + H_{grad}\left\{\boldsymbol{\nabla P}\right\} + H_{mech}\left\{\boldsymbol{P}, \boldsymbol{\varepsilon},\rho_+\right\} + H_{self}\left\{\rho_{+},\rho_{-},\rho_{\pm}\right\} \right . \nonumber \\
    &&  \left . + H_{elec}\left\{\boldsymbol{P}, \boldsymbol{E} = -\nabla \psi,\rho_{+},\rho_{-}\right\}   + H_{mix}\left\{\rho_{+},\rho_{-},\rho_{\pm},\nabla \rho_{+}, \nabla \rho_{-},\nabla \rho_{\pm}\right\} \right], \\
    &\equiv& \int d\mathbf{r} \,h\left\{\boldsymbol{P},\boldsymbol{\nabla P},\rho_{+},\rho_{-},\rho_{\pm},\nabla \rho_{+}, \nabla \rho_{-},\nabla \rho_{\pm},\boldsymbol{E},\boldsymbol{\varepsilon}\right\} \label{eq:energyfunction}
\end{eqnarray}
where $H_{LGD}$ is the Landau-Ginzburg-Devonshire (LGD) energy density\cite{rabe2007modern} and written in terms of time-dependent polarization by invoking local-equilibrium approximation\cite{groot1962non}. Similarly, $H_{grad}$ is the gradient/interfacial energy density capturing the effects of inhomogeneous polarization in the long-wavelength limit. Explicit expressions for these contributions are presented in Appendix A for  $\mbox{BaTiO}_{3}$ in a seminal work by Chen and co-workers\cite{chen2007appendix}. Coupling between the polarization and the strain is encoded in $H_{mech}$, which is the mechanical strain energy density so that $\boldsymbol{\varepsilon}$ is the strain tensor. Considering limit of small deformation, the mechanical strain energy density\cite{Landau1986} can be defined as
\begin{equation}
    H_{mech} = \frac{1}{2}\boldsymbol{\sigma}(\mathbf{r},t): \left[\boldsymbol{\varepsilon}(\mathbf{r},t) - \boldsymbol{\varepsilon^0}(\mathbf{r},t)\right] \equiv \frac{C_{ijkl}}{2} \left[\varepsilon_{ij}(\mathbf{r},t)-\varepsilon_{ij}^0(\mathbf{r},t)\right]\left[\varepsilon_{kl}(\mathbf{r},t)-\varepsilon_{kl}^0(\mathbf{r},t)\right],
\end{equation}
where $\varepsilon_{ij}(\mathbf{r},t) = \left[\partial u_i/\partial x_j + \partial u_j/\partial x_i\right]/2$ is  the $ij$ element of the total lattice-strain tensor\cite{mura1987general} so that $x_i$ are components of the spatial vector $\mathbf{r}$, $u_i(\mathbf{r},t)$ is $i^{th}$ component of the displacement vector of lattice, $\boldsymbol{\sigma}$  is the stress tensor, and $C_{ijkl}$ is the rank four elasticity tensor. Ferroelectric materials can have spontaneous strain even in stress-free conditions and such a strain tensor (eigenstrain\cite{mura1987general}) is denoted as $\varepsilon_{ij}^0$, results from electrostriction and Vegard effects, and can be defined as\cite{chen2002phase,chen2007appendix,kelley2022oxygen}, 
\begin{equation}
    \varepsilon_{ij}^0(\mathbf{r},t) = Q_{ijkl}P_k(\mathbf{r},t) P_l(\mathbf{r},t) + w_{ij}^v \rho_+(\mathbf{r},t), \label{eq:eigenstrain}
\end{equation}
where $Q_{ijkl}$ is a rank four order tensor. In this notation, $\varepsilon_{ij}(\mathbf{r},t)$ is the tensor containing both, elastic and eigenstrain, components of the strain. Furthermore, effects of the vacancies on the strain is included by the last term in Eq. ~\ref{eq:eigenstrain}, where $w_{ij}^v$ are phenomenological parameters. Here, Einstein's notation of sum over repeated indices is used.  

$H_{self}$ is the self energy density for creating the vacancies, electrons and their pairs, written as
\begin{equation}
    H_{self} = \sum_{i=+,-,\pm} G_{io} \rho_i(\mathbf{r},t) 
\end{equation}
where $G_{io}$ is the self-energy\cite{israelachvili92a} for creating $i$. $H_{elec}$ is the \textit{excess} electrical energy density written as\cite{marcus_1} 
\begin{equation}
    H_{elec} = \left[z_+ \rho_+(\mathbf{r},t) -  \rho_-(\mathbf{r},t)\right]e \psi(\mathbf{r},t) - \frac{\epsilon_0 \epsilon_{\infty}}{2} \mathbf{E}^2(\mathbf{r},t) -  \left[\mathbf{P}(\mathbf{r},t) \cdot \mathbf{E}(\mathbf{r},t)\right],
\end{equation}
where $\mathbf{E}(\mathbf{r},t) = -\nabla  \psi(\mathbf{r},t)$ is the electric field and $\psi$ is the electrostatic potential. $z_+$ is valency of oxygen vacancy and $e$ is charge of an electron. Here, the pairs of the vacancies and the electrons are assumed to carry no charge. Furthermore, $-\frac{\delta \left[\int d\mathbf{r} H_{elec}\right]}{\delta E(\mathbf{r}',t)} = \mathbf{D}(\mathbf{r}',t) = \epsilon_0 \epsilon_{\infty}\mathbf{E}(\mathbf{r}',t) + \mathbf{P}(\mathbf{r}',t)$ can be readily identified as the dielectric displacement vector.  

Excess entropy of mixing vacancies, electrons, and their pairs is defined as $H_{mix}$ along with the entropic cost of generating their inhomogeneous density profiles, written as\cite{kumar2017}
\begin{eqnarray}
    H_{mix} &=& k_B T \sum_{i=+,-,\pm}\left[\rho_i(\mathbf{r},t) \ln\left[\frac{\rho_i(\mathbf{r},t)}{\rho_{io}}\right] + \frac{1}{2}\kappa_{i}|\nabla \rho_i(\mathbf{r},t)|^2 \right] \label{eq:hmix}
\end{eqnarray}
Logarithmic terms in Eq. ~\ref{eq:hmix} can be derived by considering the number of ways in which the vacancies, the electrons, and the pairs can be distributed in space, such that their total number remain fixed. $\kappa_i$ is the coefficient of the square-gradient term\cite{kumar2014}, which penalizes inhomogeneous density profiles of $i$.

With the dissipation and energy functional given by Eqs. ~\ref{eq:dissipation} and ~\ref{eq:energyfunction}, respectively, the Rayleighian has been specified (cf. Eq. ~\ref{eq:ray}) completely with the quantities like $\tau_p, L_{kk'}, M_{kk'},$ and $\omega$ taken as inputs. After constructing the Rayleighian, a set of equations can be systematically derived by optimizing $R$ with respect to $\frac{\partial \mathbf{P}(\mathbf{r},t)}{\partial t}, \mathbf{\hat{j}}_{i=+,-}, \mathbf{\hat{j}}_{i=+,-}\cdot \mathbf{\hat{n}}$, $S$, and $\mathbf{v}_l(\mathbf{r},t)$. The set is complemented by two additional equations: one for the Lagrange's multiplier, $p$, and other one for the electrostatic potential, $\psi$. In total, nine coupled equations are derived in the next subsection. 

\subsection{Governing Equations: Linear Irreversible Thermodynamics}
We assume that the electrostatic potential adjust itself so fast that stationary condition $\frac{\delta H}{\delta \psi(\mathbf{r},t)} = 0$ is satisfied at all times and at all locations. Explicitly, this leads to 
\begin{equation}
    \epsilon_0\epsilon_{\infty}\nabla^2 \psi(\mathbf{r},t) - \nabla \cdot \mathbf{P}(\mathbf{r},t) + e\left[z_+ \rho_+(\mathbf{r},t) - \rho_-(\mathbf{r},t)\right] = 0, \label{eq:poisson}
\end{equation}
or equivalently, 
\begin{equation}
    \nabla \cdot \mathbf{D}(\mathbf{r},t) = e\left[z_+ \rho_+(\mathbf{r},t) - \rho_-(\mathbf{r},t)\right], \label{eq:poissonD}
\end{equation}
where the right hand side is the local charge density. Evaluating $\delta R(t)/\delta \left\{\frac{\partial \mathbf{P}(\mathbf{r},t)}{\partial t}\right\} = 0$, we get governing equations for the three components of polarization $P_1, P_2, P_3$ such that
\begin{equation}
    \frac{\partial \mathbf{P}(\mathbf{r},t)}{\partial t} + \left[\mathbf{v}_l(\mathbf{r},t)\cdot \nabla\right]\mathbf{P}(\mathbf{r},t) + \frac{1}{2}\mathbf{P}(\mathbf{r},t) \times \left[\nabla \times \mathbf{v}_l(\mathbf{r},t)\right] =  -\frac{1}{\tau_p} \frac{\delta \left[\int d\mathbf{r}'h\left\{\mathbf{P}(\mathbf{r}',t)\right\}\right]}{\delta \mathbf{P}(\mathbf{r},t)} \equiv \mathbf{P}^{\star}(\mathbf{r},t),\label{eq:phaseP}
\end{equation}
Functional $h$ is defined in Eq. ~\ref{eq:energyfunction} and its dependencies on variables other than the polarization are suppressed here. Similarly, $\delta R(t)/\delta \mathbf{v}_l(\mathbf{r},t) = 0$ leads to 
\begin{eqnarray}
\nabla \cdot \boldsymbol{\sigma}(\mathbf{r},t) = \frac{\tau_p}{2}\left\{\nabla \left[\mathbf{P}^{\star}(\mathbf{r},t)\cdot \mathbf{P}(\mathbf{r},t)\right] - \mathbf{P}(\mathbf{r},t)\times \left[\nabla \times \mathbf{P}^{\star}(\mathbf{r},t)\right]-\mathbf{P}^{\star}(\mathbf{r},t)\times \left[\nabla \times \mathbf{P}(\mathbf{r},t)\right]\right . &&\nonumber \\
\left . -\mathbf{P}(\mathbf{r},t)\left[\nabla \cdot \mathbf{P}^{\star}(\mathbf{r},t)\right] + \mathbf{P}^{\star}(\mathbf{r},t)\left[\nabla \cdot \mathbf{P}(\mathbf{r},t)\right]\right\}&& \nonumber \\
&& \label{eq:vlopt}
\end{eqnarray}
where we have used $d H_{mech}/d t = \boldsymbol{\sigma}(\mathbf{r},t): \left[\nabla \mathbf{v}_l(\mathbf{r},t) - \frac{\partial \varepsilon_{ij}^0(\mathbf{r},t)} {\partial t}\right]$.
With the constitutive relation,
\begin{equation}
    \sigma_{ij}(\mathbf{r},t) = C_{ijkl}\left[ \varepsilon_{kl}(\mathbf{r},t) - \varepsilon_{kl}^0(\mathbf{r},t) \right] \label{eq:stress_strain}
\end{equation}
Eq. ~\ref{eq:vlopt} can be used to study effects of strains. We should point out that in general, $\sigma_{ij}(\mathbf{r},t)$ should be computed from gradients of lattice velocity. Here, we choose a simpler and intuitive linear constitutive relation (cf. Eq. ~\ref{eq:stress_strain}) between stress and strain. Optimizations of $R(t)$ with respect to $\mathbf{\hat{j}}_{k=+,-}$ and $\mathbf{\hat{j}}_{k=+,-}\cdot \mathbf{\hat{n}}$ give 
\begin{eqnarray}
    \nabla \left [ \mu_k(\mathbf{r},t) -  \mu_{\pm}(\mathbf{r},t)\right] &=& - \sum_{k'=+,-}L_{kk'}(\mathbf{r},t)\rho_{ko}\rho_{k'o}\mathbf{\hat{j}}_{k'}(\mathbf{r},t) \quad \mbox{in the volume} \label{eq:exchange}\\
    \mu_k(\mathbf{r},t) -  \mu_{\pm}(\mathbf{r},t) &=& \sum_{k=+,-}M_{kk'}(\mathbf{r},t)\rho_{ko}\rho_{k'o}\left[\mathbf{\hat{j}}_{k'}(\mathbf{r},t)\cdot \mathbf{\hat{n}}\right] \quad \mbox{at the surface,}
\end{eqnarray}
respectively. Here, $\mu_k(\mathbf{r},t) = \delta \left[\int d\mathbf{r}'h\left\{\rho_k(\mathbf{r}',t)\right\}\right]/\delta \phi_k(\mathbf{r},t)$ and becomes the local chemical potential of $k$ in the steady state.  

Evaluating $\delta R(t)/\delta S(\mathbf{r},t) = 0$ gives 
\begin{eqnarray}
    S(\mathbf{r},t) &=& \frac{1}{\omega(\mathbf{r},t)}\left[  \frac{\mu_+(\mathbf{r},t)}{\rho_{+o}} + z_+ \frac{\mu_-(\mathbf{r},t)}{\rho_{-o}} - \frac{\mu_{\pm}(\mathbf{r},t)}{\rho_{\pm o}} + p(\mathbf{r},t)\left\{\frac{1}{\rho_{+o}} + \frac{z_+}{\rho_{-o}} - \frac{1}{\rho_{\pm o}}\right\}\right] \label{eq:reactionrate}
\end{eqnarray}
Plugging the expression for $S$ from Eq. ~\ref{eq:reactionrate} in Eq. ~\ref{eq:fluxconstraintre}, an equation for $p$ is obtained, which closes this set of equations along with the boundary conditions shown in Eqs. ~\ref{eq:rateMre} and ~\ref{eq:rateCre}.

\subsection{Non-linear Reaction Kinetics}
Eq.  ~\ref{eq:reactionrate} shows that the reaction rate depends linearly on the chemical potentials, which limits the validity of the model presented here. 
We can improve this by introducing non-linear relations between the reaction rate and the chemical potentials e.g., similar to those forming the basis of the Eyring's rate 
of reactions\cite{GLA41}. In this subsection, we present such a model, which can lead to the non-linear reaction kinetics such as in autocatalytic reactions\cite{kumar2021harnessing}. For such a purpose, 
we rewrite Eq.  ~\ref{eq:reactionrate} as a limiting case of the non-linear relation
\begin{eqnarray}
    S_{NL}(\mathbf{r},t) &=& \frac{k_B T}{\omega(\mathbf{r},t)}\left[  \exp\left[\frac{\mu_+(\mathbf{r},t) + p(\mathbf{r},t)}{\rho_{+o}k_B T} + z_+\frac{\mu_-(\mathbf{r},t) + p(\mathbf{r},t)}{\rho_{-o}k_B T}\right] \right . \nonumber \\
&& \left .  - \exp\left[\frac{\mu_{\pm}(\mathbf{r},t) + p(\mathbf{r},t)}{\rho_{\pm o}k_B T}\right]\right] \label{eq:SNL}
\end{eqnarray}
Now, the Rayleighian, $R \equiv R_{NL}$, based on $S_{NL}$ can be constructed using Eq. ~\ref{eq:ray} and it can be shown that 
\begin{eqnarray}
    R_{NL}(t) &=& R(t) + \frac{1}{2}\int d\mathbf{r}\, \omega(\mathbf{r},t) \left[S_{NL}(\mathbf{r},t) - S(\mathbf{r},t)\right]^2
\end{eqnarray}
Note that $R(t)  = -W(t) < 0$ as per the Onsager's variational principle, where the inequality is valid for a positive dissipation function and the governing equations derived in the last subsection. In contrast, $R_{NL}(t)$ can have either sign for $\omega(\mathbf{r},t) > 0$. We should point out that $1/\omega(\mathbf{r},t) = \exp\left[\frac{\mu^{\star}(\mathbf{r},t)}{\rho^{\star o}k_B T}\right] \sim \rho^{\star o}> 0$ is expected on the basis of the Eyring's rate
of reactions so that $\mu^{\star}(\mathbf{r},t)$ and $1/\rho^{\star o}$ are the chemical potential and volume of an activated complex\cite{GLA41}, respectively. Physically, this relation between the prefactor $\omega(\mathbf{r},t)$ and $\mu^{\star}(\mathbf{r},t)$ implies that the rate of reaction is linearly proportional to the concentration of the activated complexes\cite{GLA41}. More importantly, probability of realizing a kinetic path with the non-linear reaction rates is given by the Onsager-Machulp integral\cite{machulp1953,doi2019}, based on time-integral of the difference $R_{NL}(t)-R(t)$. In particular, the probability of realizing the non-linear reaction rates should be $\exp\left[-\int_0^t dt' \,\left[R_{NL}(t')-R(t')\right]/2k_B T\right]$ as per the theoretical works of Onsager and Machulp\cite{machulp1953}.  For this probability to be non-zero and significant enough, $\omega(\mathbf{r},t)$ needs to be chosen in such a manner so that the exponential of the negative of the Onsager-Machulp integral remain close to unity. In the following, we make such a choice for $\omega(\mathbf{r},t)$ and keep all of the governing equations the same except that we replace $S$ by $S_{NL}$ to capture effects of non-linear reaction kinetics in the model for the ferroelectrics developed here. 

Now, using Eqs. ~\ref{eq:denrewritten}, ~\ref{eq:exchange} and replacing $S$ with $S_{NL}$ (cf. Eq. ~\ref{eq:SNL}), 
\begin{eqnarray}
    \frac{\partial \phi_{+}(\mathbf{r},t)}{\partial t}  &=& \nabla \cdot \left[\sum_{k'=+,-}\tilde{L}_{+k'}^{-1}(\mathbf{r},t) \nabla \left [\mu_{k'} (\mathbf{r},t) - \mu_{\pm} (\mathbf{r},t)\right]\right] - \frac{1}{\rho_{+o}}S_{NL}(\mathbf{r},t) - \nabla \cdot \left[\mathbf{\phi}_+ (\mathbf{r},t)\mathbf{ v} (\mathbf{r},t)\right] \nonumber \\
&& \label{eq:phiplusfinal}\\
\frac{\partial \phi_{-}(\mathbf{r},t)}{\partial t}  &=& \nabla \cdot \left[\sum_{k'=+,-}\tilde{L}_{-k'}^{-1}(\mathbf{r},t) \nabla \left [\mu_{k'} (\mathbf{r},t) - \mu_{\pm} (\mathbf{r},t)\right]\right] - \frac{z_+}{\rho_{-o}}S_{NL}(\mathbf{r},t) - \nabla \cdot \left[\mathbf{\phi}_- (\mathbf{r},t)\mathbf{ v} (\mathbf{r},t)\right] \nonumber \\
&& \label{eq:phinegfinal}
\end{eqnarray}
where $\tilde{L}_{kk'}^{-1}$ is the $kk'$ element of the inverse matrix of $L_{kk'}\rho_{ko}\rho_{k'o}$. Three independent elements $\tilde{L}_{kk'}^{-1}$ can be interpreted\cite{newman2021electrochemical} in terms of the ionic conductivity, transference number of the vacancies related to their partial ionic currents, and diffusion constants of the vacancy-electron pairs.  

\section{Results} \label{sec:results}
In here, we present analysis of some limiting cases to highlight key effects of the vacancies and novel aspects of the model. Numerical results obtained by solving the coupled equations will be presented in a separate publication. In the model presented here, we can capture non-linear effects of barriers on reaction rates and study characteristic time for reactions. In the following, we consider three limiting cases: 1) reaction dominated regime leading to identification of a characteristic time, 2) steady state analysis highlighting coupling between the strain, electric potential, and vacancies, 3) vacancy-free regime, where a coupling between the fast and the slow component of the polarization can lead to effects of geometry manifesting in the stabilization of new topological configurations. 
 
\subsection{Reaction Dominated Regime: Characteristic time}
We consider a limiting case, where vacancies, electrons, and their pairs are homogeneously distributed so that
diffusive flux is minimal. In this limit, we need to consider $\phi_{i=+,-,\pm}(\mathbf{r},t) \equiv \phi_i^h(t)$, which satisfy (cf. Eqs. ~\ref{eq:phiplusfinal}-~\ref{eq:phinegfinal})
\begin{eqnarray}
    \frac{\partial \phi_{+}^h(t)}{\partial t}  &=& - \frac{1}{\rho_{+o}}S_{NL}^h(t) \label{eq:phiplusrxn}\\
\frac{\partial \phi_{-}^h(t)}{\partial t}  &=& - \frac{z_+}{\rho_{-o}}S_{NL}^h(t) \label{eq:phinegrxn}
\end{eqnarray}
Now, consider a case, when $\frac{1}{\rho_{\pm o}} = \frac{1}{\rho_{+o}}  + \frac{z_+}{\rho_{-o}} $ so that $\phi_{+}^h(t)+\phi_{-}^h(t) + \phi_{\pm}^h(t) = 1$ is satisfied for non-zero reaction rate, $S_{NL}(\mathbf{r},t) \equiv S_{NL}^h(t)$, given by (cf. Eq. ~\ref{eq:SNL})
\begin{eqnarray}
    S_{NL}^h(t) &=& K_0(t)\left[  \exp\left[\frac{\mu_+^h(t)}{\rho_{+o}k_B T} + z_+\frac{\mu_-^h(t)}{\rho_{-o}k_B T}\right] - \exp\left[\frac{\mu_{\pm}^h(t)}{\rho_{\pm o}k_B T}\right]\right] \label{eq:SNLrxn}
\end{eqnarray}
where we have defined $K_0(t) = \frac{k_B T}{\omega^h(t)} \exp\left[\frac{p^h(t)}{\rho_{\pm o}k_B T}\right]$ and used the notation $\omega(\mathbf{r},t) \equiv \omega^h(t), p(\mathbf{r},t) \equiv p^h(t), \mu_+(\mathbf{r},t) \equiv \mu_+^h(t), \mu_-(\mathbf{r},t) \equiv \mu_-^h(t), \mu_{\pm}(\mathbf{r},t) \equiv \mu_{\pm}^h(t)$. From Eq. ~\ref{eq:energyfunction}, we get
\begin{eqnarray}
 \frac{\mu_+^h(t)}{\rho_{+0}k_B T} &=&  \ln \phi_+^h(t) + \frac{G_{+0} + z_+ e\psi^h(t)}{k_B T} - \frac{w_{ij}^v \sigma_{ij}^h(t)}{k_B T} \label{eq:muplus}\\
\frac{\mu_-^h(t)}{\rho_{-0}k_B T} &=&  \ln \phi_-^h(t) + \frac{G_{-0} - e\psi^h(t)}{k_B T} \label{eq:muneg}\\
\frac{\mu_{\pm}^h(t)}{\rho_{\pm 0}k_B T} &=&  \ln \phi_{\pm}^h(t) + \frac{G_{\pm 0}}{k_B T} \label{eq:mupair}
\end{eqnarray}
where $\sigma_{ij}(\mathbf{r},t) \equiv \sigma_{ij}^h(t), \psi(\mathbf{r},t) \equiv \psi^h(t)$. Using Eqs. ~\ref{eq:muplus}-~\ref{eq:mupair}, we can write Eq. ~\ref{eq:SNLrxn} as 
\begin{eqnarray}
    S_{NL}^h(t) &=& K_A(t)\phi_+^h(t)\left[\phi_-^h(t)\right]^{z_+} - K_D(t)\phi_{\pm}^h(t) \label{eq:SNLhomo}\\
K_A(t) &=& K_0(t)\exp\left[\frac{G_{+0} + z_+ G_{-0}}{k_B T} - \frac{w_{ij}^v \sigma_{ij}^h(t)}{k_B T} \right] \\
K_D(t) &=& K_0(t)\exp\left[\frac{G_{\pm 0}}{k_B T}\right]
\end{eqnarray}
where $K_A(t)$ and $K_D(t)$ are defined as the time-dependent association and dissociation constants, respectively. 
It should be noted that the reaction rate $S_{NL}^h(t)$ is independent of the electrostatic potential due to the assumption of a uniform potential. In general, small variations of the electrostatic potential will lead to the dependence of the reaction rate on the potential. Furthermore, the reaction rate depends on the stress due to the Vegard's law. 
  
\begin{figure}[htb]
                \includegraphics[width=1.0\linewidth]{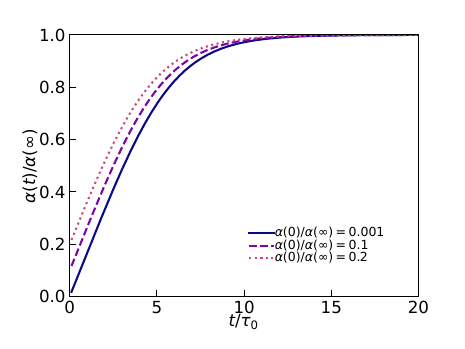}
        \caption{Solution of Eq. ~\ref{eq:alpha2} is shown here for different initial conditions.}
        \label{fig:rxn}
\end{figure}

Now, we consider the case of oxygen vacancies so that $z_+ = 2$, and assume that $K_A(t) \equiv K_{A0}$ and $K_D(t) \equiv K_{D0}$. In this case, we can solve Eqs. ~\ref{eq:phiplusrxn}, ~\ref{eq:phinegrxn}, and ~\ref{eq:SNLhomo} in terms of a time-dependent parameter, $\alpha(t)$ so that $\phi_+^h(t) = \alpha(t)/\rho_{+0}, \phi_-^h(t) = z_+ \alpha(t)/\rho_{-0}$, and $\phi_{\pm}^h(t) = 1 - (1/\rho_{+0} + z_+/\rho_{-0})\alpha(t)$ and $\alpha(t)$ satisfies
\begin{eqnarray}
    \frac{\partial \alpha(t)}{\partial t}  &=& - \frac{4K_{A0}}{\rho_{+o}\rho_{-o}^2}\left[\alpha^3(t) + \left(\frac{1}{\rho_{+0}} + \frac{2}{\rho_{-0}}\right)\frac{K_{D0}}{K_{A0}}\frac{\rho_{+o}\rho_{-o}^2}{4} \alpha(t) - \frac{K_{D0}}{K_{A0}}\frac{\rho_{+o}\rho_{-o}^2}{4} \right] \label{eq:alpha}
\end{eqnarray} 
Although Eq. ~\ref{eq:alpha} can be solved exactly but the exact solution obscure identification of a characteristic time. In here, we consider a situation so that Eq. ~\ref{eq:alpha} can be approximated as 
\begin{eqnarray}
    \frac{\partial \alpha(t)}{\partial t}  &=& - \frac{4K_{A0}}{\rho_{+o}\rho_{-o}^2}\left[\alpha^3(t) - \frac{K_{D0}}{K_{A0}}\frac{\rho_{+o}\rho_{-o}^2}{4} \right] \label{eq:alpha1}
\end{eqnarray}
Integrating Eq. ~\ref{eq:alpha1} leads to 
\begin{eqnarray}
    \ln \left[\frac{\frac{\alpha(t)}{\alpha(\infty)}-1}{\frac{\alpha(0)}{\alpha(\infty)}-1}\right]^2   - \ln \left[\frac{\left(\frac{\alpha(t)}{\alpha(\infty)}\right)^2 + \frac{\alpha(t)}{\alpha(\infty)} + 1}{\left(\frac{\alpha(0)}{\alpha(\infty)}\right)^2 + \frac{\alpha(0)}{\alpha(\infty)} + 1}\right] = - \frac{t}{\tau_0} &&\nonumber \\
+ 2\sqrt{3}\arctan\left[\frac{1}{\sqrt{3}}+ \frac{2}{\sqrt{3}}\frac{\alpha(t)}{\alpha(\infty)}\right] - 2\sqrt{3}\arctan\left[\frac{1}{\sqrt{3}}+ \frac{2}{\sqrt{3}}\frac{\alpha(0)}{\alpha(\infty)}\right]&&\label{eq:alpha2}
\end{eqnarray}
where $\tau_0$ is the characteristic time for the dissociation of a divalent oxygen vacancy-electron pair and is given by 
\begin{eqnarray}
\tau_0 &=& \frac{\rho_{+o}\rho_{-o}^2}{24K_{A0}\alpha^2(\infty)} = \frac{1}{6}\left[\frac{\rho_{+o}\rho_{-o}^2}{4}\right]^{1/3}\frac{1}{K_{A0}^{1/3}K_{D0}^{2/3}}
\end{eqnarray}
Solution of Eq. ~\ref{eq:alpha2} is plotted in Fig. ~\ref{fig:rxn}, which shows that for $t/\tau_0 > 10$, the dissociation of vacancy-electron pair is almost complete i.e., $\alpha(t)/\alpha(\infty) \rightarrow 1$. This implies that the model developed here can be simplified to some extent by considering only dissociated vacancies and electrons for capturing kinetic effects at times much greater than $\tau_0$. For example, construction of polarization-electrostatic potential loop can be constructed without considering the vacancy-electron pairs at time scales much greater than $\tau_0$, typical of experimental rates at which voltages are sweeped across ferroelectric films\cite{kelley2022oxygen}. But our complete formalism also allows us to describe ultra-fast timescales within the same framework with relative ease.
 
\subsection{Steady State Analysis: Implications of the Vegard's law}
Spatiotemporal responses of oxygen vacancies and electrons to the electrostatic potential and the strain encoded in Eqs. ~\ref{eq:phiplusfinal} and ~\ref{eq:phinegfinal} are asymmetric due to the fact that $z_+ = 2$ and the use of the Vegard's law so that only the vacancies affect the strain. In order to make this clearer, the functional derivatives (exchange chemical potentials) appearing in Eqs. ~\ref{eq:phiplusfinal} and ~\ref{eq:phinegfinal} are evaluated as

\begin{eqnarray}
    \left[\frac{\mu_+(\mathbf{r},t)}{k_B T}-\frac{\mu_{\pm}(\mathbf{r},t)}{k_B T}\right] &=& \ln \left(\frac{\phi_+(\mathbf{r},t)}{1-\phi_+(\mathbf{r},t) - \phi_-(\mathbf{r},t)} \right) + \frac{z_+ e\psi(\mathbf{r},t)}{k_B T} + \frac{G_{+o}-G_{\pm o}}{k_B T} \nonumber \\
    && - \frac{w_{ij}^v \sigma_{ij}(\mathbf{r},t)}{k_B T} - \frac{\kappa_+ \rho_{+o}}{k_B T}\nabla^2 \phi_+(\mathbf{r},t) - \frac{\kappa_{\pm o} \rho_{\pm o}}{k_B T}\nabla^2 (\phi_+(\mathbf{r},t)+\phi_-(\mathbf{r},t))\nonumber \\
&& \label{eq:vacancyflux}\\
    \left[\frac{\mu_-(\mathbf{r},t)}{k_B T}-\frac{\mu_{\pm}(\mathbf{r},t)}{k_B T}\right] &=& \ln \left(\frac{\phi_-(\mathbf{r},t)}{1-\phi_+(\mathbf{r},t) - \phi_-(\mathbf{r},t)} \right) - \frac{e\psi(\mathbf{r},t)}{k_B T} + \frac{G_{-o}-G_{\pm o}}{k_B T} \nonumber \\
    && - \frac{\kappa_+ \rho_{+o}}{k_B T}\nabla^2 \phi_+(\mathbf{r},t) - \frac{\kappa_{\pm o} \rho_{\pm o}}{k_B T}\nabla^2 (\phi_+(\mathbf{r},t)+\phi_-(\mathbf{r},t)) \label{eq:electronflux}
\end{eqnarray}
Note here that local stress-dependent term only appears in Eq. ~\ref{eq:vacancyflux} and results from our assumption that the vacancies affect the local strain, which appear in Eq. ~\ref{eq:eigenstrain}. In here, we show that the local polarization depends on spatial distribution of the vacancies and electrons via the electrostatic potential and doesn't depend solely on the electric field as in the classical ferroelectrics. First, we consider local equilibrium, which correspond to a steady state of the time-dependent equations. The local equilibrium is defined as the conditions $\left[\frac{\mu_+(\mathbf{r},t)}{k_B T}-\frac{\mu_{\pm}(\mathbf{r},t)}{k_B T}\right] = 0$ and $\left[\frac{\mu_-(\mathbf{r},t)}{k_B T}-\frac{\mu_{\pm}(\mathbf{r},t)}{k_B T}\right] = 0$.   
 
At the steady state and representing local equilibrium, $\mathbf{P}^{\star}(\mathbf{r},t)  = 0, p = \mbox{constant}$ and hence, leads to $\nabla \cdot \mathbf{\sigma}(\mathbf{r},t) = 0$. Representing all of the variables (independent of time, $t$ at $t\rightarrow \infty$) in the steady state by subscript $s$, we get $\sigma_{ij}(\mathbf{r},t) \equiv \sigma_{ij,s}(\mathbf{r}) = 0$, where the latter equality represents equilibrium (i.e., a stress-free state). This, in turn, implies that $\epsilon_{ij}(\mathbf{r},t) \equiv \epsilon_{ij,s}(\mathbf{r}) = \epsilon_{ij,s}^0(\mathbf{r}) = Q_{ijkl}P_{k,s}(\mathbf{r})P_{l,s}(\mathbf{r}) + w_{ij}^v\rho_{+o}\phi_{+,s}(\mathbf{r})$. For a weakly inhomogeneous distribution of the vacancies and the electrons so that the derivative terms in Eqs. ~\ref{eq:vacancyflux} and ~\ref{eq:electronflux} can be ignored, $\left[\frac{\mu_+(\mathbf{r},t)}{k_B T}-\frac{\mu_{\pm}(\mathbf{r},t)}{k_B T}\right] = 0$ and $\left[\frac{\mu_-(\mathbf{r},t)}{k_B T}-\frac{\mu_{\pm}(\mathbf{r},t)}{k_B T}\right] = 0$ lead to
\begin{eqnarray}
     \phi_{+,s}(\mathbf{r}) &=&  \frac{1}{1 + \exp\left[\frac{ez_+\psi_s(\mathbf{r})-w_{ij}^v \sigma_{ij,s}(\mathbf{r})}{k_B T}+ \frac{G_{+o}-G_{\pm o}}{k_B T} \right]\left[1 + \exp\left[\frac{e\left[\psi_s(\mathbf{r})-\left\{G_{-o}-G_{\pm o}\right\}\right]}{k_B T}\right]\right]} \\
    \phi_{-,s}^-(\mathbf{r}) &=&  \frac{1}{1 + \exp\left[-\frac{e\left[\psi_s(\mathbf{r})-\left\{G_{-o}-G_{\pm o}\right\}\right]}{k_B T}\right]\left[1 + \exp\left[-\frac{\left[ez_+\psi_s(\mathbf{r})-w_{ij}^v \sigma_{ij,s}(\mathbf{r})\right]}{k_B T}-\frac{G_{+o}-G_{\pm o}}{k_B T}\right]\right]}
    \label{eq:theorysol}
\end{eqnarray}
where we have used the notation $\psi(\mathbf{r},t\rightarrow \infty) = \psi_s(\mathbf{r})$ and $\sigma_{ij}(\mathbf{r},t\rightarrow \infty) = \sigma_{ij,s}(\mathbf{r})$. Using these equations, it is clear that the effects of vacancies on the total strain appear via the Vegard strain and leads to a result (valid at the steady state) 

\begin{eqnarray}
\epsilon_{ij,s}(\mathbf{r}) = Q_{ijkl}P_{k,s}(\mathbf{r})P_{l,s}(\mathbf{r}) + \frac{w_{ij}^v\rho_{+o}}{1 + \exp\left[\frac{ez_+\psi_s(\mathbf{r})-w_{ij}^v \sigma_{ij,s}(\mathbf{r})}{k_B T}+ \frac{G_{+o}-G_{\pm o}}{k_B T} \right]\left[1 + \exp\left[\frac{e\left[\psi_s(\mathbf{r})-\left\{G_{-o}-G_{\pm o}\right\}\right]}{k_B T}\right]\right]}\nonumber \\
&&
\end{eqnarray}
A similar result without any consideration of the self-energy terms ( i.e., without $G_{io}$) was derived in Ref. \cite{kelley2022oxygen}.
An additional effect of the vacancies is to affect the local electric field. At the local equilibrium, $\mathbf{P}_s(\mathbf{r}) = \alpha(\mathbf{r}) \mathbf{E}(\mathbf{r})$, where the prefactor $\alpha(\mathbf{r})$ depends on the specific form of $H$ and can be determined numerically for any functional form of $H$. As $\mathbf{E}(\mathbf{r}) = \mathbf{E}_0(\mathbf{r}) + \mathbf{E}_{1}(\mathbf{r})$, where $\nabla \cdot \left[\left\{\epsilon_0\epsilon_{\infty} + \alpha(\mathbf{r})\right\}\mathbf{E}_0(\mathbf{r})\right]  = 0$
and $\nabla \cdot \left[\left\{\epsilon_0\epsilon_{\infty} + \alpha(\mathbf{r})\right\}\mathbf{E}_{1}(\mathbf{r})\right] = e z_+ \rho_{+o}\phi_{+,s}(\mathbf{r}) - e\rho_{-o}\phi_{-,s}(\mathbf{r})$.  In other words, local electric field has contribution, $\mathbf{E}_{1}$,  resulting from inhomogeneous distribution of vacancies and oppositely charged carriers. It should be noted that $\mathbf{E}_{1}  = \mathbf{E}_{0}$ in the absence of the vacancies but $\mathbf{E}_{1}  \neq \mathbf{E}_{0}$ in the presence of the charged vacancies. This, in turn, implies that the local polarization $\mathbf{P}_s(\mathbf{r}) = \alpha(\mathbf{r}) \left[\mathbf{E}_0(\mathbf{r}) + \mathbf{E}_1(\mathbf{r})\right]$ is intimately connected with spatial distribution of vacancies and electrons. In summary, the strain-electrostatic potential loop will be asymmetric with respect to the sign of the electrostatic potential, in qualitative agreements with recent experiments\cite{kelley2022oxygen}.

\subsection{Coupling between the fast and the slow components of the polarization: topological effects in vacancy-free regime}
The model developed here is based on the decomposition of the net local polarization into a slowly-varying component, $\mathbf{P}$ and a fast component, $\mathbf{P}_e$. A coupling between these two components appear in the form of $\epsilon_{\infty}$ in the model, which affects the electrostatic potential, $\psi(\mathbf{r},t)$. In addition, another coupling appears in Eq. ~\ref{eq:phaseP} in the form of the lattice velocity, $\mathbf{v}_l$, which can be interpreted in terms of the rate of the change of the net local polarization. Molecular origin of this interpretation is the fact that local displacement of electrons and ions of ferroelectric crystals contribute towards the net polarization, which are treated in the model as $\mathbf{P}_e$ and $\mathbf{P}$, respectively. This, in turn, leads to the relation  $\mathbf{v}_l(\mathbf{r},t) \sim \partial (\mathbf{P}_e(\mathbf{r},t) + \mathbf{P}(\mathbf{r},t))/\partial t$. In the most of the phase field models, $\mathbf{v}_l(\mathbf{r},t)$ is taken to be zero, which implies that $\mathbf{v}_l(\mathbf{r},t)\simeq  \partial (\mathbf{P}_e(\mathbf{r},t))/\partial t \rightarrow 0$ at the time-scales relevant to the models. This implies that Eq. ~\ref{eq:phaseP} can be written as
\begin{equation}
    \frac{\partial \mathbf{P}(\mathbf{r},t)}{\partial t} = \mathbf{P}^{\star}(\mathbf{r},t),\label{eq:phaseP1}
\end{equation}
where Eqs. ~\ref{eq:energyfunction} and ~\ref{eq:phaseP} allow us to identify
\begin{eqnarray}
    \mathbf{P}^{\star}(\mathbf{r},t) &=& -\frac{1}{\tau_p}\left[\frac{\delta \left[\int d\mathbf{r}'\,H_{LGD}\left\{\mathbf{P}(\mathbf{r}',t)\right\} + H_{grad}\left\{\nabla \mathbf{P}(\mathbf{r}',t)\right\}\right]}{\delta \mathbf{P}(\mathbf{r},t)} - \mathbf{E}(\mathbf{r},t))\right] \label{eq:pstar}
\end{eqnarray}
In general, $H_{LGD} + H_{grad}$ can be written in powers of $\mathbf{P}$ so that 
\begin{eqnarray}
    H_{LGD}\left\{\mathbf{P}\right\} + H_{grad}\left\{\nabla \mathbf{P}\right\} &=& \frac{1}{2\chi}\mathbf{P}^2(\mathbf{r},t) + 
\frac{\kappa_{p,1}}{2}\left[\nabla \cdot \mathbf{P}(\mathbf{r},t)\right]^2 +
\frac{\kappa_{p,2}}{2}\left[\nabla \times \mathbf{P}(\mathbf{r},t)\right]^2 \nonumber \\
&& +  \kappa_{p,3}\mathbf{P}(\mathbf{r},t) \cdot \left[ \nabla \times \mathbf{P}(\mathbf{r},t)\right] \label{eq:hlgd2}
\end{eqnarray}
Eqs. ~\ref{eq:pstar} and ~\ref{eq:hlgd2} lead to 
\begin{eqnarray}
    \mathbf{P}^{\star}(\mathbf{r},t) &=& -\frac{1}{\tau_p}\left[\frac{1}{\chi}\mathbf{P}(\mathbf{r},t) - \kappa_{p,1} \nabla\left[\nabla \cdot \mathbf{P}(\mathbf{r},t)\right] + \kappa_{p,2} \nabla \times \left[\nabla \times \mathbf{P}(\mathbf{r},t)\right] + 2 \kappa_{p,3} \left[ \nabla \times \mathbf{P}(\mathbf{r},t)\right] \right . \nonumber \\
&& \left . - \mathbf{E}(\mathbf{r},t)\right],\label{eq:phasePsimplified}
\end{eqnarray}
Using Eq. ~\ref{eq:poisson} in the absence of the vacancies and the electrons; and $\nabla \times\mathbf{E}(\mathbf{r},t) = 0$, 
\begin{eqnarray}
    \nabla \cdot \mathbf{P}^{\star}(\mathbf{r},t) &=& -\frac{1}{\tau_p}\left[\left\{\frac{1}{\chi} + \frac{1}{\epsilon_0 \epsilon_{\infty}}\right\}\nabla \cdot \mathbf{P}(\mathbf{r},t) - \kappa_{p,1} \nabla^2\left[\nabla \cdot \mathbf{P}(\mathbf{r},t)\right] \right]\label{eq:divPstar}\\
\nabla \times \mathbf{P}^{\star}(\mathbf{r},t) &=& -\frac{1}{\tau_p}\left[\frac{1}{\chi}\nabla \times \mathbf{P}(\mathbf{r},t) + \kappa_{p,2} \nabla \times \left(\nabla \times\left[\nabla \times \mathbf{P}(\mathbf{r},t)\right]\right) + 2 \kappa_{p,3}\nabla \times\left[\nabla \times \mathbf{P}(\mathbf{r},t)\right] \right] \nonumber \\
&& \label{eq:curlPstar}
\end{eqnarray}
Operating with divergence and curl on Eq. ~\ref{eq:phaseP1}, and using Eqs. ~\ref{eq:divPstar}, ~\ref{eq:curlPstar}, we get
\begin{eqnarray}
    \frac{\partial \left[\nabla \cdot\mathbf{P}(\mathbf{r},t)\right]}{\partial t} &=& \left[\frac{\kappa_{p,1}}{\tau_p} \nabla^2 - \frac{1}{\tau_L} \right] \left[\nabla \cdot\mathbf{P}(\mathbf{r},t)\right] \\
\frac{\partial \left[\nabla \times \mathbf{P}(\mathbf{r},t)\right]}{\partial t} &=& \left[\frac{\kappa_{p,2}}{\tau_p} \nabla^2 -  \frac{2 \kappa_{p,3}}{\tau_p}\left(\nabla \times\right) - \frac{1}{\tau_p \chi} \right] \left[\nabla \times \mathbf{P}(\mathbf{r},t)\right]
\end{eqnarray}
where $\tau_L = \tau_p/\left\{\frac{1}{\chi} + \frac{1}{\epsilon_0 \epsilon_{\infty}}\right\}$ is the characteristic time for the change of $\nabla \cdot\mathbf{P}(\mathbf{r},t)$ and $\tau_p \chi$ is the chracteristic time for the change of $\nabla \times \mathbf{P}(\mathbf{r},t)$. Note that $\tau_L/(\tau_p \chi) = 1/(1 + \chi/(\epsilon_0 \epsilon_{\infty})) \ll 1$ for $\chi\gg 1$. This means that for $t \gg \tau_L$, $\nabla \cdot \mathbf{P}(\mathbf{r},t) = 0$. This allows us to identify relation between the local polarization and the electric field by using Eq. ~\ref{eq:phaseP1}. For example, in a steady state, $\mathbf{P}^{\star}(\mathbf{r},t) = 0$, which leads to a relation between the local polarization and the local electric field as 
\begin{eqnarray}
\frac{1}{\chi}\mathbf{P}_s(\mathbf{r}) - \kappa_{p,1} \nabla\left[\nabla \cdot \mathbf{P}_s(\mathbf{r})\right] + \kappa_{p,2} \nabla \times \left[\nabla \times \mathbf{P}_s(\mathbf{r})\right] + 2 \kappa_{p,3} \left[ \nabla \times \mathbf{P}_s(\mathbf{r})\right] 
&=& \mathbf{E}_s(\mathbf{r}),\label{eq:phasePsteady}
\end{eqnarray}
where we have used the notation $\mathbf{P}_s(\mathbf{r}) = \mathbf{P}(\mathbf{r},\infty)$ and $\mathbf{E}_s(\mathbf{r}) = \mathbf{E}(\mathbf{r},\infty)$. For a given volume, solution of Eq. ~\ref{eq:phasePsteady} depends on the geometry and boundary conditions. A particular set of solutions, which enforces $\nabla \cdot \mathbf{P}_s(\mathbf{r}) = 0$ everywhere in space including the boundaries, will be discussed here. In particular, $\nabla \times \mathbf{P}_s(\mathbf{r}) = \lambda \mathbf{P}_s(\mathbf{r})$, which enforces $\nabla \cdot \mathbf{P}_s(\mathbf{r}) = 0$ for a constant $\lambda$ will be discussed here. For such a divergence-free polarization vector, Eq. ~\ref{eq:phasePsteady} demands 
\begin{eqnarray}
\left[\frac{1}{\chi} + \kappa_{p,2} \lambda^2 + 2 \kappa_{p,3} \lambda\right]\mathbf{P}_s(\mathbf{r}) &=& \mathbf{E}_s(\mathbf{r}),\label{eq:PErelation}
\end{eqnarray}
$\lambda$ can be obtained by solving for an eigenvalue of the equation $\nabla \times \mathbf{P}_s(\mathbf{r}) = \lambda \mathbf{P}_s(\mathbf{r})$ with boundary conditions. Nevertheless, Eq. ~\ref{eq:PErelation} shows that $\mathbf{P}_s(\mathbf{r})$ and $\mathbf{E}_s(\mathbf{r})$ are parallel to each other at the steady state, which may not be true in general. In fact, Eq. ~\ref{eq:phaseP1} can be used to determine the time-dependent electric field required for setting up known polarization vector in space and time. 

In contrast to the above analysis, if we consider $\mathbf{v}_l(\mathbf{r},t) = \beta \partial \mathbf{P}(\mathbf{r},t)/\partial t, \beta$ being a constant, then Eq. ~\ref{eq:phaseP} can be written as
\begin{equation}
    \frac{\partial \mathbf{P}(\mathbf{r},t)}{\partial t} + \beta \left[\frac{\partial \mathbf{P}(\mathbf{r},t)}{\partial t}\cdot \nabla\right]\mathbf{P}(\mathbf{r},t) + \frac{\beta}{2}\mathbf{P}(\mathbf{r},t) \times \frac{\partial \left[\nabla \times \mathbf{P}(\mathbf{r},t)\right]}{\partial t} =  \mathbf{P}^{\star}(\mathbf{r},t),\label{eq:phaseP2}
\end{equation}
Eq. ~\ref{eq:phaseP2} shows that the coupling between the slow and the fast component of the net polarization will affect spatiotemporal distribution of $\mathbf{P}(\mathbf{r},t)$, while the steady state behavior of $\mathbf{P}(\mathbf{r},t) \equiv \mathbf{P}_s(\mathbf{r})$ remains the same. Eq. ~\ref{eq:phaseP2} opens up a way to study effects of confinement on the spatiotemporal distribution of topologically non-trivial polarization, which may not be realized at an equilibrium.

\section{Conclusions}~\label{sec:conclusions}
We developed a thermodynamically consistent time-dependent model for understanding the effects of multivalent vacancies on relations among polarization-electric potential and strain-electric potential in thin films of ferroelectrics. In contrast to the most of the phase field models, non-linear effects of the reaction kinetics leading to generation of charged vacancies and electrons from their pairs are introduced in the model. In addition, diffusion and elastic effects of the charged vacancies are considered, which are shown to exhibit asymmetric reponses of the strain to the electric potential. Furthermore, the model introduces coupling between the slow component and a fast component of the net polarization, which is expected to affect time-dependent relation between the polarization and the electric field. Impedance response from the thin films of ferroelectrics with vacancies should exhibit defect dipole behavior\cite{ding_influence_2018, kelley2022oxygen, akbarian2019understanding, DHAKANE2023100264} which will dynamically couple with the polar matrix giving rise to frequency-dependent behavior. In the presence of electrodes, one can also expect electrode polarization phenomenon\cite{kumar2017}  due to localization of the vacancies near an oppositely charged electrode. These localized vacancy induced  polarization phenomenon can be used to extract diffusion constant of the vacancies. A simplified model\cite{kumar2017} developed using the Rayleighian approach has been used previously to fit the impedance spectra as a function of frequency and temperature for thin films of ionic polymers. We envision that in future, experimental results for impedance spectra from thin films of ferroelectrics with vacancies can be fitted using the model developed here and extract the diffusion constant of the vacancies. The Rayleighian approach to build the model is shown to be general enough for constructing non-linear reaction kinetics.   
Application of the model to simulate motion of domain walls in ferroelectrics in the presence of vacancies will be presented in a forthcoming publication. 

 \section{Acknowledgements}
This work was supported by the Center for Nanophase Materials Sciences, which is a US DOE, Office of Science User Facility at Oak Ridge National Laboratory. 

\section{Data Availability Statement}
The data that support the findings of this study are available from the corresponding author upon reasonable request.

\bibliographystyle{unsrt}
\bibliography{ferroelectrics}

\end{document}